# Code Coverage and Test Automation: State of the Art

Karl Meinke[1]
School of Electrical Engineering and Computer Science,
KTH Royal Institute of Technology, Stockholm, Sweden.

## Introduction

This chapter surveys the state of the art in *code coverage* from the perspective of test automation. Our aim is to describe and motivate the three most popular classes of glass box *test coverage models*, which are: *control flow*, *logic* and *data flow coverage*. We take a fairly rigorous approach to code coverage models. Thus, for each class, we will give precise definitions of specific examples, some of which are widely known while others deserve to be better known by test engineers. Our main goal is to present coverage models that represent the state of the art. These should stimulate thought regarding best practice, and indicate future directions for test process improvement.

We will also discuss the connections between code coverage models and test automation. By defining coverage models in a rigorous way (usually in terms of concepts from *graph theory* and *logic*) we open the door to both algorithmic *measurement* of coverage and algorithmic *generation* of test cases, also known as *automated test case generation* (ATCG). These are two important subjects in the very broad field of test automation. Coverage measurement methods will not be discussed in this Chapter, while automated test case generation will be discussed in the last section of this Chapter.

Code coverage attempts to answer the following question:

> "*how much testing has been performed by executing a given test suite T?*",

in an objective, and measurable way. Clearly, some test suites do achieve better results than others, often at the cost of extra effort and time. An analogy from physics might be the abstract concept of "*energy*" as a unifying measure of different kinds of mechanical work. Coverage models attempt to give an objective and scientifically motivated answer to the question above. Naïve approaches, such as simply measuring the size of *T*, can be very misleading – what if all tests in *T* were very similar? A large test suite, when designed badly, can give very low code coverage. This may reflect a high degree of repetition or *redundancy* in *T*, which is an important concept that we will discuss.

---

The code coverage given by *T* is not the same as the *number of errors* detected by *T*. This is termed the *effectiveness* of *T*. Effectiveness depends upon the nature of the errors in any specific software code, aka. the *system under test* SUT, that *T* is applied to. However, we might reasonably expect that increasing the code coverage of *T* should increase the test effectiveness of *T*, all other factors being equal. Nor is coverage the same as the *effort* to construct *T*. This could be the manual effort of writing each test case by hand. It could otherwise be the time taken to generate *T* using some automated test case generation tool. However, in general increased coverage requires more effort, for both manual and automated test case construction. This effort is often a limiting factor for deploying sophisticated code coverage models.

Test coverage is sometimes used as a proxy for measuring *software reliability*, in the case that no errors have been found. This application of coverage, though it seems to be intuitively reasonable, is difficult to motivate in a scientifically legitimate way, i.e. the relationship between coverage and reliability is unclear.

Code coverage measures are most frequently encountered in the context variously known as *glass box, white box or structural testing*. In this context, we can identify and count the different structural features of the SUT, e.g. code lines, Boolean tests, assignments, etc. Outside this context, for example in black box testing, it can be quite difficult to define precise coverage measures for testing. However, since glass-box testing is widely used in industry, this is not a problem in practice.

Historically, the oldest known model of code coverage is "line coverage" which simply counts the number of lines of code exercised by executing all the tests in *T*. This measurement usually requires a *code profiling tool* which can identify the execution path in the SUT taken by running a single test case. Line coverage can be a somewhat misleading measure, since it is not invariant to code syntax or even to different programming styles. Nowadays much better graph and logic-based coverage models exist. One purpose of this Chapter is to present these models. Unfortunately, the more sophisticated a coverage measure is, the less widely it tends to be used in industry, unless it becomes legally mandated by some safety standard such as DoD 178[2]. This can also be because tool support is missing.

Why is code coverage needed? In general, it exists as a measure of both the quality of the SUT (an artifact) and the quality of the testing process (an activity). Code that passes a high coverage test suite could be considered high quality code. While a test team that consistently generates and uses high coverage test suites could be considered a very proficient team.

However, there are also more technical roles for coverage. One important application of coverage is as a *stopping criterion* for testing: we can stop when we reach 100% code coverage under some agreed model (such as line coverage). Another use, already mentioned above, is to estimate the code reliability when no errors have been found by a test suite *T*. This is probably a controversial use, since it is difficult to establish any rigorous scientific link between code coverage and code reliability. We are always in the dark about whether errors actually exist in the SUT that have simply not been found by *T*. This issue borders on the more general subject of *software reliability*, which has a much broader scope than software testing. We can only point to the literature of this extensive subject e.g. (Lyu, 1996).

---

[2] This was the fate of the complex MCDC coverage model.

Just as we can compare different test suites for the coverage they provide, we can also compare the coverage models themselves. We might naively look for the "best" coverage model, and only use that. From a theoretical viewpoint, we can indeed compare two coverage models in a *relative* way. We will present a common technical approach later. Such comparison helps the test engineer to choose an appropriate coverage model for a specific situation. Technically, we may find that one coverage model *subsumes* another. If coverage model *A* subsumes model *B* then *A* is always (i.e. on every SUT) at least as good as *B*, and usually better. An absolute or *quantitative comparison* of coverage models *A* and *B* is more difficult. However, we can compare them empirically, over a large set of SUT examples, to gain some insight into their absolute performance. Here we face the problem that no universally agreed standardized set of SUT benchmarks for this comparison exists. So, comparison results are easily skewed significantly by the data set. This means it can be difficult to be confident about choosing the "best" coverage model for a particular testing task. Some degree of uncertainty will always remain and experimentation will always be necessary.

## Scope of this Chapter.

This Chapter is based on lecture notes used for eight years in the course DD2459 *Software Reliability* taught at KTH Royal Institute of Technology in Stockholm. The course syllabus and structure were originally inspired by the book: "*Introduction to Software Testing*" by Amman and Offut (Ammann & Offutt, 2017), which was the recommended course textbook. The reader is encouraged to consult this book, where more technical depth about coverage models is needed. On the other hand, in this Chapter we also discuss state-of-the-art test automation issues such as test case generation. Such issues rarely appear in textbooks on testing. Perhaps this is due to the level of mathematical difficulty of the problem.

The intended audience of this Chapter is managers, developers and testing professionals who need to revise or update their knowledge of the subject, as well as university level students meeting this subject for the first time. We assume that the reader has some experience of code development, and an appreciation for the syntactic features of typical programming languages. Some knowledge of *graph* *theory* will helpful, which can be found in many books on discrete mathematics e.g. (Biggs, 2003). But essentially, the Chapter is self-contained.

## Coverage and Test Automation.

Why should we study code coverage specifically in the context of test automation? There are at least two reasons.

1. (Basic reason) Code coverage models can (and should) be precisely defined so that coverage can be quickly, accurately and reliably measured, preferably in a tool vendor independent way. Then coverage measurement can be automated as one of several important test automation services.

2. (Advanced reason) Code coverage models can be used to define both the individual test requirements and the corresponding test case data needed to achieve 100% coverage. As we will show: combining a coverage model *A* with a specific SUT *P* generates a precise set *TR* of test requirements, which can be represented as mathematical constraints (for example to take a specific execution path through *P*). Such constraints can be solved by hand, or possibly algorithmically, leading to tools for test case generation. A solution to a test requirement $tr \in TR$ is therefore concrete test case data (e.g. an input vector) *tc* that can be executed on *P*. Nowadays, the existence of efficient constraint solving algorithms, means that some test cases can be quickly and automatically generated, by solving *tr*. This avoids tedious and error prone manual construction. There are even commercial tools offering this functionality, though the field is not yet mature. Unfortunately, there are also strong theoretical and fundamental limitations in this area[3]. Current commercial solutions for automated test case generation are mostly restricted to the simple views of code provided by model-based testing. But there are some notable successes. Furthermore, advanced machine learning (ML) algorithms are beginning to make inroads into the field of automated test case generation (Meinke, Bennaceur, & Hähnle, 2018). So, this is indeed the "state of the art" for test automation.

## Glass box Testing

By *glass box testing* we mean any software testing that is carried out from the perspective of the structure of the code that is to be tested. The latter is variously known as the *system under test*, *software under test* (SUT), *unit under test* (UUT) *application under test* (AUT) or *implementation under test* (IUT). For this reason, glass box testing is also known as *white box* or *structural testing*, where the last title is perhaps the most appropriate and descriptive.

---

[3] The first limitation is the *unsolvability of the halting problem* (Sipser, 2012).

By visually inspecting the code structure, we can decide what code features to execute during a test, and hence how to design the test cases themselves. An automated syntactic analysis of the code could also give such information. It may also be possible to predict what the intended outcome of a test case should be, in terms of observable behavior or output. At the very least, we can assume that the SUT should not crash, i.e. terminate prematurely due to an *untrapped exception*.

The technical idea underlying glass box testing can be simply stated.

> *An implementation error in code (if it exists) must exist in some syntactic location: (e.g. a line number, Boolean expression, assignment command, numerical expression etc.). If that location is never executed by any test case then the error has no chance to manifest itself in an observable way.*

So, the general aim of glass box testing is to identify and exercise code locations in systematic ways that can avoid missing errors by oversight.

**Definition 1**. Glass box or structural testing is the process of exercising software with test cases defined by syntactic features of the source code or system under test.

What Definition 1 rules out is test cases based on the software API or end user requirements. These are also valid test requirements. They are just not glass box requirements.

In fact, Definition 1 is a little too simply stated. In reality, what matters is not just executing some syntactic code feature, but the *way* that we execute it. This is because even though there are *multiple paths* through any real-world program, there are also distinctively *different ways* to execute the *same path*. If our goal is to actively search for errors, then we need to be both systematic and creative to uncover them. In practice then, test coverage models are related to complex *path properties* of code rather than just static syntactic features. The concept of a code path will be precisely defined later.

This shift from syntactic code features to code paths quickly leads to the heart of the testing problem, which is the size of the test suite to be constructed and executed.

A so-called *straight-line program*, i.e. a program without any branches, loops or recursion, has just one path for execution. A *branching program* (using only `if-then-else` and/or `case` statements) has a tree structure and hence a path set which grows exponentially in the number of Boolean conditions. Still the number of paths is finite, and execution of all code paths is technically possible. Programs in the real world however generally have complex nested loops and/or recursion. For such programs, the number of paths is infinite. To execute all possible paths with a finite test suite $T$ is impossible. Instead, some simplification is required to select a finite subset of paths from among the infinite possibilities by using a coverage model. Even with such simplifications, the size of *T* tends to grow *exponentially* with the size of the SUT. For this reason, we rarely see glass box methods used outside the context of *unit testing* – i.e. testing the individual code units or models, such as classes and methods.

## Glass Box Coverage

Usually glass box testing has the goal to exercise a minimum set of locations or paths. In an age of continuous integration (CI)/continuous development (CD) the commercial emphasis is often on getting code delivered as fast as possible. So, from a business perspective, the less time spent on software testing the better. This goal conflicts with another desirable business goal, which is to maintain high product quality and hence high customer satisfaction. Therefore, the software tester is commonly under pressure to find a small number of highly effective test cases, since software tests may take several minutes or even hours to execute. Thus, test selection becomes an optimization problem that may be difficult and time consuming to solve optimally. Furthermore, test selection begs the whole question of whereabouts and what kind of SUT errors we are likely to find.

Against this commercial background, the philosophy of glass box coverage is to be *systematic* in testing, on the grounds that we simply don't know where errors are likely to be found. This approach is at odds with *experience based* or *human-centric* approaches to software testing that might try to identify "typical errors" that programmers make, and test for these directly. In our view, if human-centric approaches are the "art" of software testing, then coverage-based approaches are the "science" of testing. Probably there is a need for both types of approach. In any case, the scientific approach is easier to communicate and study.

Software testers every day face the question "*how much testing is enough*"? This question is:
   (i)     an economic problem (when does the budget run out),
   (ii)    a scheduling problem (when will testing finish) and
   (iii)   a technical problem (which test cases still need to be written).

The coverage-based approach to testing gives a simple answer to these problems. *Testing can stop when we reach 100% coverage under an agreed coverage model*. If the model is precise and measurable then termination can be precise identified. In practice, testing may terminate when only 90% coverage is reached, or some other acceptable figure, for pragmatic reasons. Whatever the situation, one role of coverage is to define for the testing team "*this much testing is enough*".

We have already suggested that coverage models come about because the *size of the test suite is an unreliable indicator of the quality of testing*. To take an extreme case, consider a test suite $T$ that consists of 1000 tests. Ostensibly $T$ is 10 times better than a much smaller subset $S \subseteq T$ consisting of just 100 tests. In fact, all 1000 test cases of $T$ execute the same single code path $p$. In this case, the additional 900 test cases do not add to the knowledge of $S$ outside of $p$.

Most glass box coverage models are agnostic about statistical usage patterns of the code: all usage patterns are equally likely. An alternative is to randomly select a subset of glass box tests according to a known statistical usage pattern. This approach borders on the fields of *statistical testing*, *risk-based testing*, *random testing* and *reliability testing*. However, by systematically evaluating the SUT under a neutral view of usage, we can build up a more comprehensive and broadly-based view of its reliability that hopefully includes some "black swan" events.

## Glass Box versus Black box Testing

Lest the reader fall into the error of believing that glass box testing and glass box coverage are the only possible technical approaches, we should briefly consider at least one major alternative.

In *black box testing*. the code structure is hidden and test cases must be designed from some other perspective, typically a user-centric or developer–centric perspective. This means test requirements originate either in:
    (i)     the code API or,
    (ii)    end-user/developer requirements.

We will not go into the technical details here, as black box testing is beyond the scope of this Chapter. Suffice it to say that each approach (glass box and black box) has its strengths and weaknesses. Furthermore, a strength of one approach may be a weakness of the other. For example, glass box testing cannot cope with a "sin of omission" i.e. an unimplemented requirement. In this case there is literally no feature or path to test! Moreover, glass box coverage makes no statement about the code being fit for purpose, since code structure has no specific relationship with user requirements. On the other hand, black box and requirements-based testing methods make no statements about unintended functionality (such as a Trojan horse or a security issue). Thus, we can see that these two methods are complementary, and both are important for software quality assessment.

## The Three Classes of Glass Box Coverage Models

Following (Ammann & Offutt, 2017), we will classify coverage models into three different families. This is mainly a pedagogic device to help the tester compare different coverage models, and understand how they attempt to improve upon one another. These three classes are:

1. *Control Flow Models*

2. *Logic Models*

3. *Data Flow Models*

*Control flow models* specify the various execution paths that test cases should invoke during testing. Examples of control flow models include: *line coverage*, *node coverage* and *edge coverage*. The idea is to exercise at least the "main" paths through the code. These models are among the oldest and most widely used. They are easy to unambiguously define and measure, and therefore generally quite widely used in industry. However, they offer a rather minimal solution to the problem of finding errors.

*Logic coverage models* attempt to identify and analyze the influence of key Boolean expressions in the SUT that determine the flow of control. They can therefore be viewed as a powerful extension of control flow models. Logic coverage even analyzes critical combinations of components or sub-formulas in Boolean expressions. The subtle point here is that different sub-formulas can lead to executing the same code path, but in profoundly different ways. These execution differences often relate well to the programmer's design intention (or design misunderstanding!). Examples include *branch coverage*, *predicate coverage*, *clause coverage* and the *MCDC* coverage model mandated by the avionic safety standard DO178C. Logic coverage models are a modern approach to testing, and represent the state of the art. Coverage tools are sometimes lacking (at least as open source tools) and it will clearly take the industry some time to adopt these models widely.

The final category is *data flow models*. These are orthogonal to control flow models. Whereas a control flow test might identify errors in control flow structure (e.g. a bad loop guard), a data flow test could identify errors in data flow, (e.g. a bad assignment statement). Data flow coverage raises issues around the *data types* used in a program, and whether these are correctly implemented and used. Data flow coverage models are not yet widely used in industry, but probably deserve to be better known.

## Distinguishing Test Requirements from Test Cases

Before we begin our survey of the above three classes of coverage models, it is necessary to step back and take a more abstract view of testing. In particular, we need to define and distinguish the related concepts of:
    (i)    a *test requirement* and,
    (ii)   a *test case*.

Every glass box coverage model defined in the literature characterizes a set of code features (e.g. paths) to be executed by testing. However, such models cannot identify the *specific* input data (i.e. the concrete test cases themselves) that would lead to executing these features. The reason is obvious; the coverage model definition cannot know about the specifics of *your* particular SUT, which was unknown to the authors of the model.

Therefore, we must separate two concepts:

(i) *what* must be executed (i.e. the *test requirements*, e.g. specific paths), and

(ii) *how* shall these locations be executed (i.e. the *test cases* as specific concrete data values with a certain representation).

The test engineer should be able to derive test cases from a clear and specific set of test requirements. It should be possible to unambiguously show that a test case *fulfills* a test requirement to which it can be *traced*. Ideally, every test requirement is *implemented* by at least one test case. It should be possible to identify a *complete* set of requirements in some sense, and hence to establish that the *test suite* (i.e. the set of all concrete test cases) is also complete. All these methodological goals can be satisfied using the coverage concepts and models we present in this Chapter.

Since a glass box coverage model only defines the test requirements, the test engineer is left to puzzle out the concrete test cases that implement these. In fact, the problem of *test case construction*, can be highly non-trivial, as students in software testing courses quickly find out. Manual test case construction is notorious for being slow and error prone, and often every bit as difficult as coding itself. Furthermore, a test construction process that is error prone can hardly be expected to compensate for bad coding!

Therefore, *automated test case generation* (ATCG) from test requirements has been an important long-term goal in the field of test automation. The idea is to devise an algorithm that can automatically generate a test case that fulfills a given test requirement. Since a test requirement can often be represented as a mathematical constraint, and since constraint solving algorithms exist, this idea is not so far-fetched. However, there are a number of fundamental theoretical reasons why ATCG is technically difficult. Indeed, some theoretical results show that no efficient and general-purpose ATCG algorithms will ever be found. Nevertheless, the approach is interesting and useful algorithms have been found and will be developed in the future. Besides the subject of constraint solving, the subject of machine learning (ML) is also starting to target this goal (Meinke, Bennaceur, & Hähnle, 2018).

## Test Redundancy

The separation of abstract test requirements from concrete test cases allows us to formulate important concepts of *redundancy* in testing. Here, the word "redundancy" essentially means performing the same test task twice (or more!).

One aspect of test redundancy is when we implement the same test requirement with more than one test case. The problem is complicated by the fact that a test case can implement more than one test requirement at the same time. Ideally, a test suite should have low redundancy (preferably zero) to save time, energy, effort of test writing, verdict construction etc. However, as an optimization problem, zero-redundancy test suites are not always easily achieved. Again, there are fundamental theoretical hurdles. Furthermore, redundancy usually exists in *both* the test requirements and the test cases themselves, but for different reasons.

Let us consider some simple examples of redundancy. Redundancy exists between two test requirements *A* and *B* when every test case that satisfies or fulfills requirement *A*, necessarily satisfies requirement *B* as well. Consider the following (line numbered) code fragment:

```
line n:         y = y+1;
line n+1:       z = f(y);
```

According to the model of *line coverage*, this code has two test requirements:

$LC_1$: *execute line n*
$LC_2$: *execute line n+1*

The requirement $LC_1$ is redundant because of the usual interpretation of the semicolon construct ";" in most structured programming languages. If line `n+1` is executed it is only because line `n` was executed and terminated successfully[4]. So, *any* test case that satisfies requirement $LC_2$ will also satisfy $LC_1$. It is unnecessary to consider two separate requirements here. Unfortunately, while some redundancy amongst test requirements can often be quickly identified by visual inspection, complete test requirement optimization is usually impossible. This is because of the requirement that *every* test case that fulfills requirement *A* also fulfills requirement *B*. This statement is quantified over infinitely many test cases, so a "proof" of redundancy tends to be either trivial (as above) or very hard!

Even when redundancy is reduced at the test requirement level, it can still persist on the level of concrete test cases. As we have just seen, a single test case can satisfy several requirements, e.g. every test for requirement $LC_2$ is a test for $LC_1$ above.

---

[4] Obviously you should check if your programming language conforms to this interpretation of the symbol ";".

However, a single test requirement can also be satisfied or fulfilled by more than one test case. If we see *satisfaction* as a binary relation between test requirements and test cases, then we are asserting here that satisfaction is a *many-to-many relation*, as shown in Figure 1 Coverage Models, Test Requirements and Test Cases .

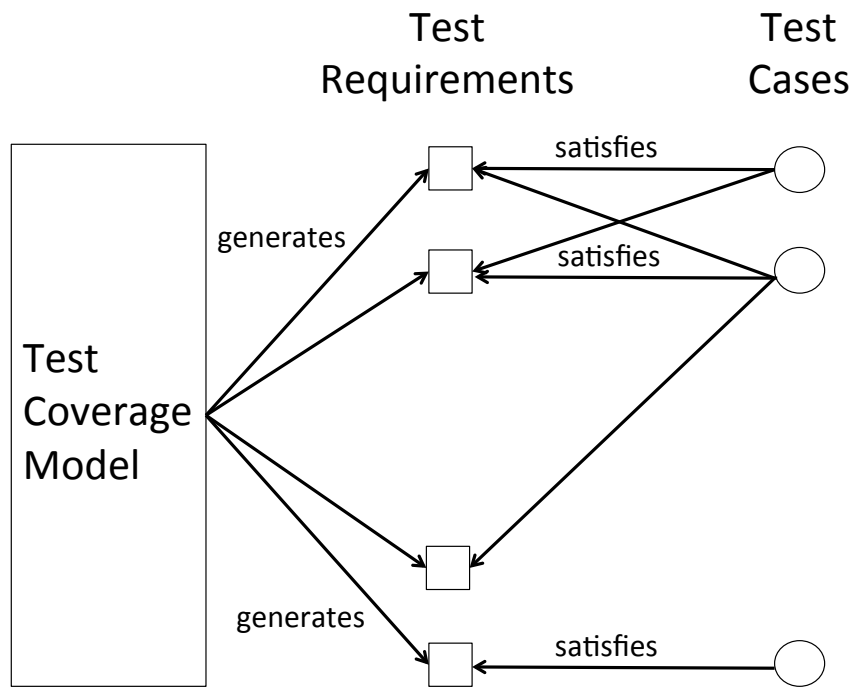

Figure 1 Coverage Models, Test Requirements and Test Cases

## Control Flow Coverage Models

In this Section we will survey some of the basic control flow coverage models that have been proposed by the testing community. To provide a unified discussion of all three classes of coverage models, we begin by considering computer programs as *graphs*. Graphs are a powerful way to abstract away from the particular features of specific programming languages to discover general principles for software testing.

### Condensation Graphs

A graph representation of a program provides a two-dimensional, visual representation that can capture both syntactic and semantic information about the code. The semantic information is very important for testing is it helps to characterize the program behavior (at least in schematic terms). To capture this semantic information, a graph must be constructed with reference to the precise syntax and semantics of the underlying programming language.

The *parse tree* (or *syntax graph* or *syntax tree*) for a program is well known in compiler theory. It typically just captures the *scope* of the program's constructs, where this scope corresponds to a specific sub-tree.

A *control flow graph* (CFG) for a program is more complex than a parse tree and has a more general (not necessarily tree-like) structure. The aim is to capture the execution control flow or runtime sequencing of the individual program instructions. We will always assume that a program *p* (the SUT) is syntactically correct. Otherwise it may be ambiguous what the CFG for *p* should even be. A CGF for a program *p* is a directed labeled graph *G* with nodes labeled by the atomic instructions of *p*. The cycles and branches of G must reflect the semantics of the different programming language constructs, e.g. loops, if_then_else statements, etc. If the `goto` construct is allowed in a programming language then its CFGs can have any graph theoretic structure at all! For this reason, most programmers eschew `goto` or use it sparingly. As CFGs, simple *tree-like graphs* are only rarely seen, corresponding to straight line code or simple types of branching programs. In general, branching programs coincide with *directed acyclic graphs* (DAGs), which are slightly more general than trees.

From a testing perspective, a CFG is a useful *model* of code, but this model can have too much information. For test coverage, we can introduce a more abstract (i.e. less detailed) model of code, which is the *condensation graph*.

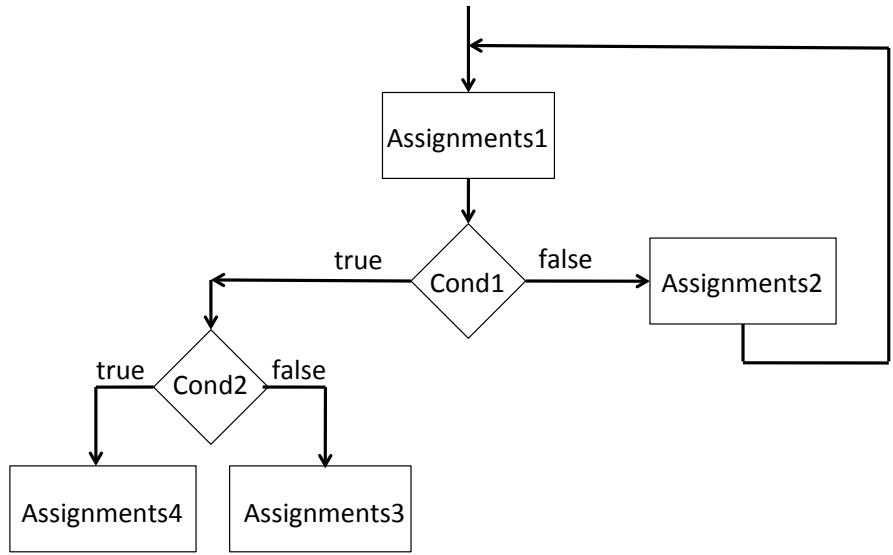

**Figure 2 Schematic condensation graph**

Figure 2 Schematic condensation graph shows a typical example of condensation graph structure. A condensation graph is a *labeled directed graph* (see Figure 5: Basic Graph Theory [i] below) with two kinds of nodes: *condition nodes* represented as diamonds, and *assignment nodes* represented as rectangles. Each condition node is labeled by a Boolean condition `cond` or guard coming from a program control flow statement such as:

```
if (cond) S
if (cond) S₁ else S₂
while (cond) S
case cond₁: S₁; cond₂: s₂; …
repeat S until(cond)
```

An assignment node is labeled by a *semicolon separated list of assignment statements*:

```
v₁ = exp₁;
v₂ = exp₂;
…
vₙ = expₙ;
```

This list is also termed an *assignment block*. From our discussion of test redundancy above, we have seen that all the assignments in the same assignment block are *indistinguishable*, i.e. to test one line in the block is to test all lines. This leads us to regard the *n* assignment statements as members of the same *equivalence class* or *partition*, a concept borrowed from discrete mathematics (Biggs, 2003). In fact, other types of statements besides assignment can also be included in an assignment block, including *function calls* `exp`$_2$`;` and *return statements* `return exp;` The key characteristic for membership is that only the *semicolon* or *sequencing* construct is used to separate the block member instructions.

Notice that the right hand side expression `exp`$_i$ in

```
v_i = exp_i;
```

can be arbitrary. So, if the expression `exp`$_i$ involves the function or procedure under definition, i.e. a *recursive function call* then a cyclic execution path is contained *implicitly* within the assignment block. This causes some difficulties when the test coverage of code loops needs to be addressed, but we will ignore this problem here. Unfolding a recursive definition is one way to deal with test coverage of recursive code.

Notice also that a condensation graph has just a single *entry node* (marked graphically by an arrow or edge "from nowhere"), but possibly multiple *exit nodes*. So, all executions start unambiguously from the same entry point.

Figure 3 Simple Pseudo Code below gives an example of some pseudo-code and Figure 4 Condensation Graph shows the corresponding condensation graph. For test requirement definition, it is generally useful to numerically index each graph node as shown in Figure 4.

```
1. x' = x;
2. y' = y;
2. while (x' != 0 & y' != 0) do {
3.   if (x' < 0) then x' = x' + 1 else x' = x' - 1;
4.   if (y' < 0) then y' = y' + 1 else y' = y' - 1;
6. }
7. if (x' == 0) then return y else return x;
```
**Figure 3 Simple Pseudo Code**

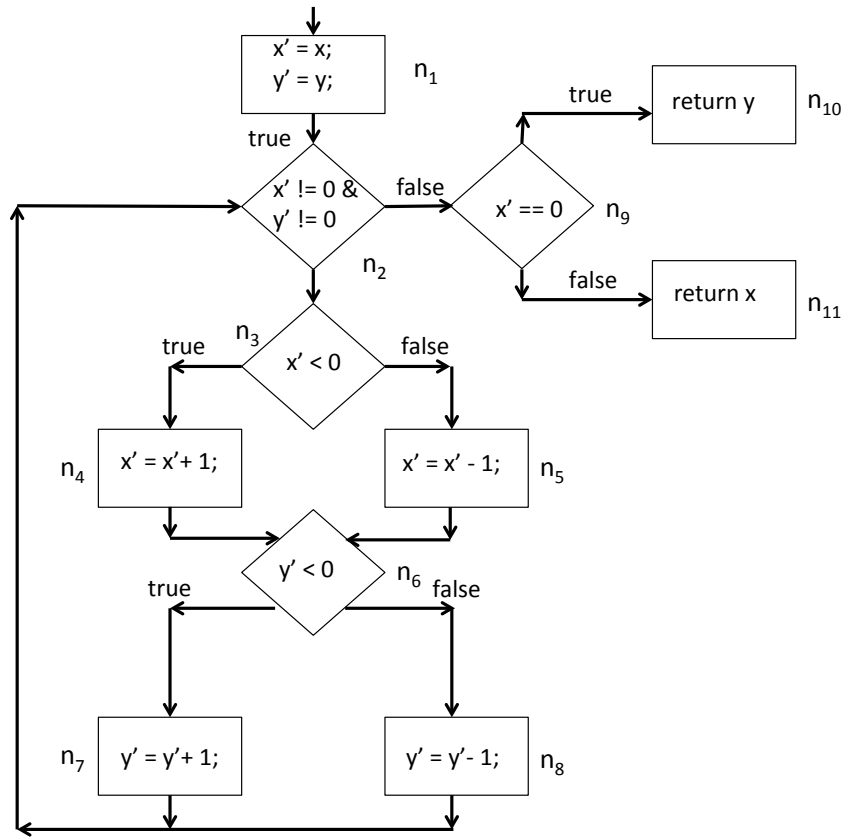

**Figure 4 Condensation Graph**

To correctly construct a set of control flow test requirements by hand (without using any tool) it is important to be able to accurately draw or at least mentally visualize the condensation graph of the code to be tested. This requires some understanding of the specific programming language syntax and semantics, as well as a general understanding of condensation graph representation. There are several tools providing this functionality, but they tend to be language specific, e.g. Soot (Vallée-Rai, o.a., 1999) for Java.

## Test Requirements for Control Flow Coverage

For control flow coverage, a test requirement is a path $p = v_1,..., v_n$ in a condensation graph $G$. The meaning of $p$ as a test requirement is that any test case $t$ satisfying $p$ must traverse all the nodes $v_1,..., v_n$ in that order.

Notice that if $t$ satisfies $p$ then $t$ satisfies any requirement that is a *sub-path* of $p$. This includes all the individual nodes in $p$. So, we can simplify $p$ by writing it as concisely as possible. Furthermore, since a concrete test case $t$ will usually traverse many nodes in a condensation graph $G$ then there are many possible requirements that $t$ can satisfy at the same time. This is the phenomenon of test redundancy that was mentioned previously.

---

**Basic Graph Theory**

Recall that a *labeled directed graph* is a pair $G = (V, E, L, \lambda: V \to L)$ consisting of a set $V = \{ v_1, ..., v_n \}$ of *vertices* or *nodes*, a set $E \subseteq V \times V$ of *edges* $(v_i, v_j)$, a set $L$ of *labels* and a *labeling function* $\lambda: V \to L$. A *path p* in a graph $G$ is a finite sequence or list of nodes

$$p = (v_1, v_2, ..., v_n)$$

such that for each adjacent pair of nodes $v_i, v_{i+1}$ in $p$, there is an edge $(v_i, v_{i+1}) \in E$. The *length* of $p$ is $n$. Notice that a path of length 1 is allowed. This consists of just a single node. A *sub-path* of $p$ is any continuous sub-sequence of $(v_1, v_2, ..., v_n)$, e.g. $(v_2, v_3)$ is a sub-path of $p$ but $(v_2, v_4)$ is not.

We let *Paths*$(G)$ denote the set of all paths in G.

We say that $p = (v_1, v_2, ..., v_n)$ is a *cycle* if $v_1 = v_n$. $G$ is *acyclic* if there is no path in $G$ that is a cycle.

---

**Figure 5: Basic Graph Theory**

## Control Flow Coverage Models

Below we give examples of proposed and often fairly well-known control flow coverage models from the testing literature.

**Definition 2**. **Node Coverage (NC):** each reachable path $p = v_1$ of length 1 in $G$ (i.e. $v_1 \in V$) is a test requirement, $p \in \text{TR}_{\text{NC}}(G)$.

Node coverage is the modern form of *line coverage*. Condensation into assignment blocks has already taken care of the test redundancy arising from lines that are only separated by the sequencing (i.e. semicolon ";") construct.

In Definition 2, *reachability* means that there exists at least one test input that actually leads program execution to $v_1$. Otherwise the code at $v_1$ (an assignment block or a Boolean condition) can be regarded as *dead* or *unreachable code*. Identification of dead code is an important achievement in itself, and usually implies a coding error of some kind in the SUT. In fact, writing glass box test cases for code coverage is one of the few ways we have to identify dead code.

Node coverage is probably the minimum acceptable level of coverage. It is worth remembering Myers' comment in his classic work (Myers, 1979) on testing: "*line coverage (NC) is so weak that it is generally considered useless*".

**Definition 2**. **Edge Coverage (EC):** Each reachable path $p = v_1, v_2$ of length $2$ in $G$ is a test requirement $p \in TR_{EC}(G)$.

Figure 6 Node coverage versus edge coverage compares EC with NC in terms of coverage.

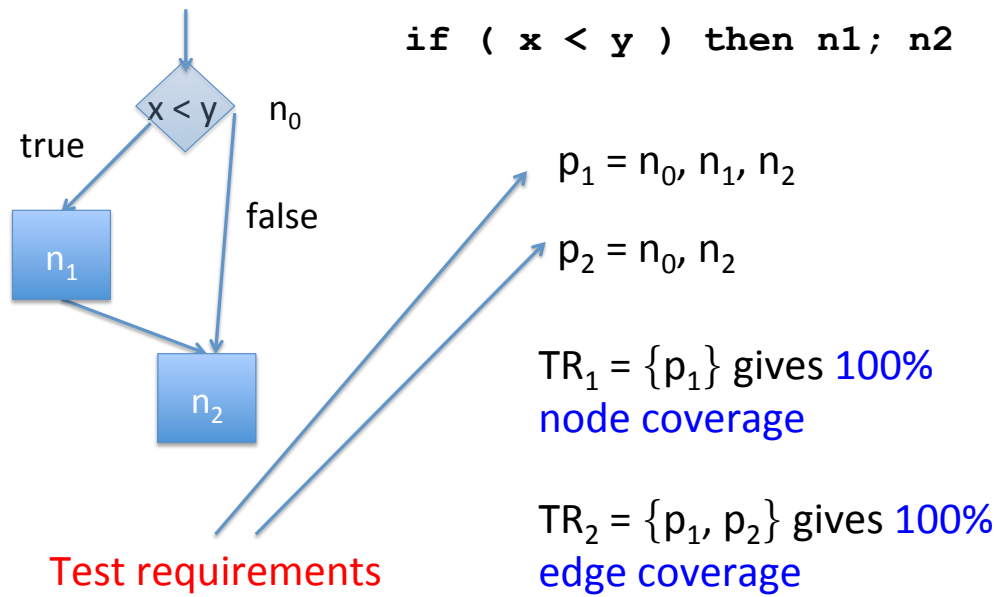

Figure 6 Node coverage versus edge coverage

In Figure 6, the test requirement set $TR_1 = \{p_1\}$ will give 100% node coverage (NC) but not 100% edge coverage (EC), as the edge ($n_0$, $n_2$) is not covered. The test requirement set $TR_2 = \{p_1, p_2\}$ gives both 100% node coverage (NC) and 100% edge coverage (EC). Not surprisingly therefore, $TR_2$ has more requirements than $TR_1$.

We can easily see that the situation in Figure 5 holds true generally.

**Fact 1**. *Any test suite TS for G that satisfies full (i.e. 100%) edge coverage must satisfy full (100%) node coverage.*

Graph theory not only allows us to rigorously define coverage models. It also allows us to *analyze* the relationships between these models. We can use the formal definitions of NC and EC to give a rigorous mathematical proof of Fact 1 as follows.

**Proof of Fact 1.** Let $G$ be any condensation graph. Let $TS$ be a test suite that satisfies 100% EC. Consider any *reachable* node $v$ in $G$ then $v \in TR_{NC}(G)$. We must show that some test case $t \in TS$ satisfies $v$ as a test requirement.

Now if $v$ is reachable then by definition, $v$ must be reachable by some path $p = v_1, v_2, \ldots, v_n$ in $G$ starting from the unique entry node of $G$, and where $v = v_n$. Then $v$ will be included in the edge requirement for $G$ consisting of just the last edge[5] $(v_{n-1}, v_n)$ in $p$, which is a reachable edge. Since $TS$ gives 100% EC for $G$ there is at least one test case $t \in TS$ that satisfies $(v_{n-1}, v_n)$, and so $t$ satisfies $v_n = v$. Since $v$ was arbitrarily chosen, then every reachable node $v$ in $G$ is satisfied by some $t \in TS$. Therefore, $TS$ satisfies 100% NC.

Going one step further from NC and EC we have:

**Definition 3**. **Edge-pair Coverage (EC²):** Each reachable path $p = v_1, v_2, v_3$ of length 3 in $G$ is a test requirement $p \in TR_{EC^2}(G)$.

Figure 7 Edge versus edge pair coverage gives a comparison of edge coverage (EC) with edge pair coverage (EC²).

---

[5] Technically here we are assuming that there exists at least one edge in G. This will be true for all but the degenerate case of a single node.

```
if ( x < y ) then n1 else n2;
if ( w < y ) then n4 else n5;
```

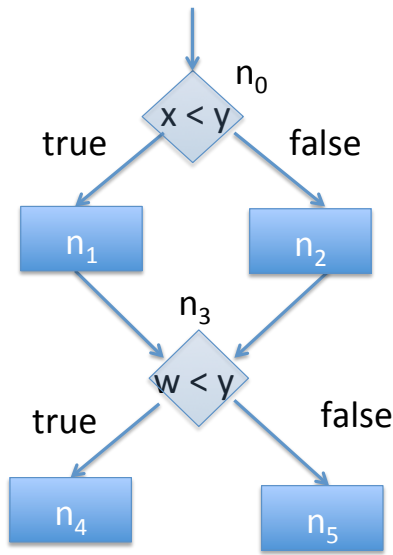

$p_1 = n_0, n_1, n_3, n_4$

$p_2 = n_0, n_2, n_3, n_5$

$p_3 = n_0, n_2, n_3, n_4$

$p_4 = n_0, n_1, n_3, n_5$

$TR_1 = \{p_1, p_2\}$ gives 100% edge coverage

$TR_2 = \{p_1, p_2, p_3, p_4\}$ gives 100% edge-pair coverage

Figure 7 Edge versus edge pair coverage

Again, we can establish that any set of test requirements that gives full edge pair coverage $EC^2$ must also give full edge coverage EC, by a similar argument to the proof of Fact 1 above. A detailed proof is left as an exercise for the reader.

Clearly we can extend Definition 3 to **$EC^n$** coverage models for $n$ = 3, 4, … . However, this can result in a combinatorial explosion in the size of the test requirement set (in the worst case). On the other hand, choosing any specific value of $n$ does not deal with the important problem of *loop coverage*, since a loop body may have unbounded length.

Let us look at test coverage for loops, arising from programming constructs such as `while <cond> do <body>` or `repeat <body> until <cond>`. A loop can be arbitrarily complex from a graph theoretic perspective since loops may be nested to any depth. Furthermore, each loop body has a finite but unbounded length.

**Definition 3: Simple Paths**. A path $p = v_1, ..., v_n$ in a condensation graph $G$ is said to be *simple* if $p$ has no repetitions of nodes other than possibly the first $v_1$ and last, $v_n$. A simple path $p = v_1, ..., v_n$ in $G$ is said to be *maximal* or *prime* if there is no path $q$ in $G$ that includes $p$ and is also simple.

So a simple path has no internal cycles, but may itself be a cycle. Unfortunately, there are usually too many simple paths in a condensation graph $G$, and therefore much redundancy among these as a set of test requirements. Maximal simple paths (which subsume all the non-maximal paths from a test perspective) remove some redundancy however.

**Definition 4: Prime Path Coverage (PPC)** Each reachable prime path $p = v_1,..., v_n$ in $G$ is a test requirement $p \in TR_{PPC}(G)$.

Figure 8 Prime Paths Example shows the set of all prime paths for a small condensation graph $G$ with a single loop.

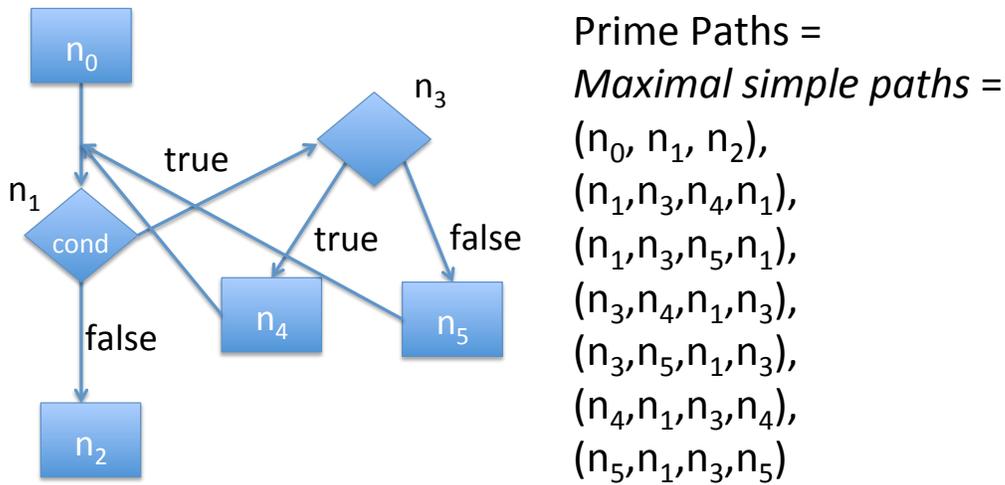

Prime Paths =
*Maximal simple paths* =
$(n_0, n_1, n_2)$,
$(n_1, n_3, n_4, n_1)$,
$(n_1, n_3, n_5, n_1)$,
$(n_3, n_4, n_1, n_3)$,
$(n_3, n_5, n_1, n_3)$,
$(n_4, n_1, n_3, n_4)$,
$(n_5, n_1, n_3, n_5)$

**Figure 8 Prime Paths Example**

For this example, consider the following paths in $G$:

$p_1 = (n_0, n_1, n_2)$,   $p_2 = (n_0, n_1, n_3, n_4, n_1, n_2)$,   $p_3 = (n_0, n_1, n_3, n_5, n_1, n_2)$,
$p_4 = (n_0, n_1, n_3, n_4, n_1, n_3, n_4, n_1, n_2)$,   $p_5 = (n_0, n_1, n_3, n_5, n_1, n_3, n_5, n_1, n_2)$.

Then $TR_1 = \{p_1, p_4, p_5\}$ gives 100% PPC, but $TR_2 = \{p_1, p_2, p_3\}$ does not give 100% PPC. Notice that this coverage model executes the loop body zero times and two times, but not once. Another common coverage model for loops is to execute each loop body zero times and one time. According to (Ammann & Offutt, 2017) one motivation for the PPC model is that the set of all prime paths in a condensation graph can be computed efficiently by a simple dynamic programming algorithm.

Let us conclude this section with two further coverage models.

**Definition 5: Complete Path Coverage (CPC)** Every reachable path $p = v_1, ..., v_n$ in $G$ is a test requirement, $p \in \text{TR}_{\text{CPC}}(G)$.

This coverage model is the strongest possible, in the sense that it contains every other control flow coverage model. However, if $G$ has any cycles then $\text{TR}_{\text{CPC}}(G)$ is an infinite set of test requirements, and not practical.

**Definition 6: Specified Path Coverage (SPC)** Every reachable path $p = v_1, ..., v_n$ in a set $S \subseteq \textit{Paths}(G)$ of test paths is contained in some path $p \in \text{TR}_{\text{SPC}}(G)$.

In SPC the collection of paths $S$ is supplied as a parameter. As an example of SPC, suppose that $S$ contains paths that traverse every loop free path $p$ in $G$ and every loop in $G$ exactly once. This gives a more precise definition of the commonly used coverage heuristic mentioned previously.

## Logic Coverage

Logic coverage models can be viewed as a refinement of the control flow coverage models presented in the previous section. In logic coverage we are not only interested in paths, but also in *different ways* of executing these paths, so that more subtle errors can be found[6]. This means that the test suites generated by logic coverage models are often more effective at finding errors than control flow coverage. This is at the expense of being larger and with longer execution times.

As we saw in the previous Section, a control flow test requirement is a path $p = v_1, ..., v_n$. In a condensation graph $G$, a path is actually a sequence of condition nodes and assignment blocks, with the property that two linearly adjacent assignment blocks are never allowed. Therefore, if a path $p = v_1, ..., v_n$ in $G$ has more than 1 node, then it must have at least one condition block $v_i$ (a diamond in the graphical representation above).

Looking more closely at the Boolean condition attached to a condition block $v_i$ this is a Boolean formula `cond`$_i$ which can either be an *atomic* or a *compound formula*.

An atomic formula is built up from expressions or terms `exp`$_i$ (usually featuring variables `x`, `y` appearing in the program) using common relational operators (such as `<=`, `==`) also known simply as *relations*. Also permissible in `cond`$_i$ are user defined Boolean valued relations, e.g. `Boolean my_relation(x`$_1$ `, ..., x`$_n$`)`. A compound Boolean formula is built up from atomic formulas using the common Boolean operations[7] *conjunction* (`and`), *disjunction* (`or`) and *negation* (`not`).

---

[6] There are typically many examples of such errors, such as division by zero.
[7] The choice and syntax of Boolean operators may vary between programming languages, but *sequential* or *eager* and, or and not are a common core.

Furthermore, in many programming languages the familiar Boolean operations come in an *eager* form (e.g. in Java "`&`" and "`|`") and a *lazy* form (e.g. in Java "`&&`" and "`||`"), where the latter lazy form is often important to express recursive programs.

A moment's reflection shows that when a condition node labeled by a compound formula occurs along a path *p*, then it is often possible *to take the same path p in semantically different ways.* By this we mean that different initial variable assignments will take program execution along the same path *p*, but at the condition node different Boolean sub-formulas can be true (or false). For example, given the compound formula

$$c = a \mid b$$

We could make *c* true by making just *a true*, or just *b true*, or possibly both if they are not mutually exclusive conditions.

This idea will become clearer when we look at specific logic coverage models. For now, it is enough to appreciate that since the programmer usually has a specific intension with the choice and structure of each compound Boolean formula `cond`$_i$, we should test whether this intension is correct, i.e. the Boolean formula `cond`$_i$ correctly implements the appropriate mathematical predicate at its location. For example: Boolean formulas often specify *boundary conditions* for different algorithmic steps such as loops. So, we should check whether these boundaries have been correctly represented in the code.

## Clauses and Predicates

We follow the terminology[8] of (Ammann & Offutt, 2017) to discuss the structural components of Boolean conditions in programs.

> **Definition 7**
> 1. A *clause* (aka. an atomic formula) is a Boolean valued expression with no Boolean valued sub-expression. Examples: `myBooleanVar, myGuard, x==y, x<=y, x>y`.
> 2. A *predicate* (aka. a compound formula) is a Boolean combination of clauses using (either eager or lazy) Boolean operators. For concreteness, we will use just the eager operation symbols: `&, |, !`. We also allow brackets (,) for disambiguation.
>    Examples: `p&q, (myGuard1 | myGuard2), !x==y`

---

[8] The reader should be aware that this terminology does not fully agree with the standard terminology used in logic and discrete mathematics texts.

3. We let $P(G)$ denote the *multiset[9] of all predicates* used as labels in a condensation graph $G$.
4. Let $p$ be a predicate. We let $C(p)$ denote the set of all clauses occurring in $p$. If $P$ is a set of predicates then $C(P)$ is the multiset of all clauses of all predicates of $P$.

Notice that a multiset is used in Definition 7 Part (3), since a predicate $p$ can occur as the label of different nodes $v_i$, $v_j$ in the same condensation graph $G$. We need to test each such occurrence of predicate $p$, as these will lie along different paths. The same issue applies to clauses in Part (4).

## Logic Coverage Requirements

As we saw previously, a control flow test requirement is a path. However, for logic coverage a test requirement *tr* is a logical constraint on variable values *at a specific point in code execution*. To derive a concrete test case *t* that satisfies *tr* is therefore just as difficult as for control flow coverage. A logical coverage requirement usually needs to be reformulated as a constraint on the input variables at the start of program execution, irrespective of where the clause or predicate to be tested lies. This is necessary because we cannot simply jump into the middle of the SUT to start program execution. Execution can only begin at the unique entry point of a condensation graph.

This observation makes explicit a problem that we briefly mentioned previously, the problem of *dead code* (also known as *unreachable code*). By dead code we mean a code feature (at some location) for which there is no test case at all that can execute that feature. For example, since a logic coverage requirement is a logical constraint, this constraint may not be solvable[10]. In that case we have successfully identified dead code.

Unsolvable constraints include obvious logical contradictions such as:

$$x \neq x, \quad x > x, \quad x == x+1.$$

However, much more subtle examples exist, and in general these can be hard to identify simply by visual inspection. Unfortunately, there is no universal algorithm[11]

---

[9] Recall a *multiset* is a set where each element has a *multiplicity*. For example, {a, b, a} and {a, b} are the same as sets, but differ as multisets.

[10] A logical formula $\phi$ which has no positive solutions is known as a *contradiction*. If the negation of $\phi$ has no positive solutions then $\phi$ is known as a *tautology*.

[11] Mathematical examples of this abound. For example, a *Diophantine equation* is a polynomial equation, possibly in many variables, such that only integer solutions are sought. Such equations are easily coded in most programming languages by using integer variables. The famous theorem of Y. Matiyasevic, see (Sipser, 2012),

to decide whether a constraint is solvable or not. Furthermore, even where Boolean constraints *are* known to be algorithmically solvable, finding their solutions can be computationally very expensive. The so called *SAT problem* (see e.g. (Sipser, 2012)) of finding satisfying assignments to an arbitrary Boolean formula in *n* Boolean variables is a famous example of an NP complete problem (ref), meaning that it is solvable in exponential time (as a function of *n*) but not in polynomial time, unless P = NP. This conjecture is considered unlikely by most computer scientists.

In practice, the difficulty of finding solutions to Boolean constraints varies considerably according to the structure of the SUT. The problem tends to become harder when user-defined predicates appear in a constraint, i.e. when we use a method `myPredicate(…)` that returns a `Boolean` result.

## Basic Logic Coverage Models

In this section we will look at some well-known logic coverage models. The definitions will become increasingly complicated and subtle, but the models are more powerful for exactly this reason. In theory, a sophisticated coverage model should provide more effective testing (i.e. better fault-finding capacity) than a simple model, as it takes more execution factors into account. But in our experience, the choice of the concrete test suite chosen to implement the test requirement set also greatly influences the final testing results. This choice is outside the coverage model definition itself. Therefore, we encourage the reader to acquire direct practical experience with each model.

**Definition 8**. **Predicate Coverage (PC)** For each predicate $p \in P(G)$, there are two test requirements: (1) a requirement $tr(p)_{\text{true}} \in \text{TR}_{\text{PC}}$ that implies $p$ is reached and evaluates to *true*, and (2) a requirement $tr(p)_{\text{false}} \in \text{TR}_{\text{PC}}$ that implies $p$ is reached and evaluates to *false*.

In Definition 8, reachability only concerns graph reachability, and the existence of lazy Boolean operators does not play a role (unlike clause coverage below). Since logic coverage models can look rather abstract at first sight, we will consider a running example.

**Example:** Consider the following simple condensation graph

---

asserts that it is undecidable whether a Diophantine equation has any integer solutions.

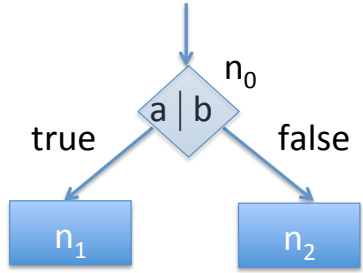

where $n_0$ is labeled by the predicate $p = a \mid b$, and $a$ and $b$ are Boolean variables of the SUT for $G$. Then $tr(p)_{true} = a \mid b$ and $tr(p)_{false} = !(a \mid b) \equiv !a \mathbin{\&} !b$ (using De Morgan's law).

Let us define two concrete test cases $tc(p)_{true}$ and $tc(p)_{false}$ which satisfy the requirements $tr(p)_{true}$ and $tr(p)_{false}$ respectively.

$tc(p)_{true} = (a = true, b = false)$
$tc(p)_{false} = (a = false, b = false)$

Notice in the above test suite that we never test for the situation $b = true$. This may well represent an important usage situation, for otherwise why would $b$ have been included in $p$?

**Definition 9: Clause Coverage (CC)** For each predicate $p \in P(G)$, and each clause $c \in C(p)$ there are two test requirements: (1) a requirement $tr(c)_{true} \in TR_{CC}$ that implies $c$ is reached and evaluates to true, and (2) a requirement $tr(c)_{false} \in TR_{CC}$ that implies $c$ is reached and evaluates to false.

Notice now in Definition 9 that the existence of lazy Boolean operators affects the reachability concept. To reach a clause we must first reach the predicate containing it (graph reachability) and then reach the clause itself (under the clause evaluation order). If a preceding clause has a constant Boolean value (true or false) a subsequent clause can be unreachable, depending upon the Boolean operators connecting them.

**Example.** Consider the same condensation graph as before and the predicate $p = a \mid b$ labeling $n_0$. Then $tr(a)_{true} = a$ and $tr(a)_{false} = !a$. Also $tr(b)_{true} = b$ and $tr(b)_{false} = !b$.

Let us define four test cases $tc(a)_{true}$ and $tc(a)_{false}$, $tc(b)_{true}$ and $tc(b)_{false}$ which satisfy the above four test requirements

$tc(a)_{true} = (a = true, b = false)$
$tc(a)_{false} = (a = false, b = true)$
$tc(b)_{true} = tc(a)_{false}$
$tc(b)_{false} = tc(a)_{true}$

Notice that in all these four test cases the predicate $p$ is always true, so we achieve 100% clause coverage (CC) but not 100% predicate coverage (PC). This means that PC and CC are *independent* coverage criteria. Neither implies the other, i.e. we can have full coverage of one model without full coverage of the other.

Since both CC and PC test important but independent aspects of code execution, it is natural seek a third coverage model that combines these two aspects. First, we shall consider a naïve or brute force solution.

**Definition 10. Combinatorial Coverage (CoC)** For each predicate $p \in P(G)$, and for every possible Boolean assignment $\alpha: C(p) \rightarrow \{true, false\}$ to all the clauses $c \in C(p)$ there is a test requirement $tr(p)_\alpha \in TR_{CoC}$ which implies $p$ is reached and the clauses of $p$ evaluate to $\alpha$.

Clearly CoC implies both PC and CC by the statement "*every possible Boolean assignment* $\alpha: C(p) \rightarrow \{true, false\}$". This encompasses all the values the clauses can take and hence all the values the predicate $p$ can take. Sometimes CoC is called *multiple condition coverage* (MCC).

The set of requirements for CoC grows linearly in the length of a path $p$ in $G$, but exponentially in the number of clauses in each predicate $p$. In practice, a predicate usually has only a small number of component clauses. However, for efficiency, we should consider whether a smaller set of test requirements might not be both adequate and effective. For this we need to consider the semantic role of clauses in predicates rather than just their syntactic role. We will discuss this later in this Chapter.

## Distributive versus Non-Distributive Coverage

At this point we should clarify a subtle technical issue regarding the interpretation of the clause coverage model and its subsequent refinements below. There are two different interpretations of clause coverage, called the *distributive* and *non-distributive* interpretations.

To explain them, let us consider a generic path $p$ of length $n$ in a condensation graph $G$:

$$p = A_1, C_2, A_3, ...., C_{n-1}, A_n$$

Without much loss of generality, we can assume that p consists of alternating assignment nodes $A_i$ and condition nodes $C_i$ that are labeled by predicates $p_i$.

For clause coverage, every condition node $C_i$ is associated with some number of test requirements $tr_1, ..., tr_k$ on the clauses $c_1, ..., c_n \in C(p_i)$ that all yield the same outcome for the predicate $p_i$. A later condition $C_j$ for $j > i$ has its own separate set of test requirements $tr'_1, ..., tr'_m$ say.

One question is: do we combine the requirements *in conjunction* (or *multiplicatively*) as a test requirement set

$$tr_1 \& tr'_1, tr_1 \& tr'_2, ...., tr_k \& tr'_m$$

or do we take their union *in disjunction* (or *additively*) as a test requirement set

$$tr_1, ..., tr_k, tr'_1, ..., tr'_m$$

Both these interpretations are legitimate from a testing perspective. However, the definition of CC does not say which interpretation to apply and they lead to very different sets of test requirements. The multiplicative interpretation is also termed *distributive* (after the Distributive Law of Boolean algebra, which it reflects) while the additive interpretation is also termed *non-distributive*.

The distributive interpretation will test a much larger collection of code executions, and therefore test the SUT more thoroughly. However, the distributive test requirement set grows exponentially in the length $n$ of $p$ (by repeated "multiplication") and this is not scalable for large $n$. So, in practice, the non-distributive approach, which grows linearly with path size $n$, is preferred.

## Advanced Logic Coverage Models

We concluded this Section by revisiting the problem of combining CC with PC. As we have observed 100% clause coverage (CC) does not imply 100% predicate coverage (PC) and vice versa. How can we combine the advantages of each in a concise and scalable way? There have been a number of proposed solutions to this problem in the testing community.

## Active Clause Coverage

Consider again our running example of the predicate $p = a \mid b$ and the test cases

$tc_1 = (a = true, b = true)$,
$tc_2 = (a = false, b = false)$.

Then the test suite $TS = \{tc_1, tc_2\}$ achieves both 100% PC and 100% CC.

Now in the disjunction $p = a \mid b$, clauses *a* and *b* have some independence from one another. More precisely, $b = false$ completely *exposes* the effect of *a* (i.e. p is fully determined by a). Similarly, for $a = false$. Conversely, $b = true$ *masks* the effect of *a* (i.e. *a* has no effect on the value of *p*). Similarly, for $a = true$.

Despite these obvious facts, the effect of clause *a* on its own (with *b* held constantly *false*) and clause *b* on its own (with *a* constantly *false*) are never tested by *TS*. Can we find something *more expressive than PC or CC* but *less expensive than CoC* which addresses independence? We will use notion of an *active clause*, which is one that determines the overall predicate value.

**Definition 11**. Given a clause *c* in a predicate *p* we say that *c determines p* under a Boolean assignment α to all the clauses of *p* if, and only if, changing the value of *c* under α (and only this value) changes the value of *p*.

Intuitively, the idea is that when *c* determines *p* then *c* has "complete control" over *p*. Such situations allow us to test each clause in isolation, and hence to query their role against the programmer's original intention. Furthermore, the number of test requirements grows linearly in the number of clauses and the size of each predicate. So, this approach is scalable. Notice that determination as defined above is a *local property* of the SUT that depends only on the truth table for *p*, and is independent of how *p* is written down (e.g. bracketing, operator order). As a local property, we analyze the truth table for *p* while ignoring its wider context in the SUT, for better or worse!

**Example.** Figure 9 Determinacy Table for *P* shows the truth table for a predicate *P* with three clauses, C, D and E.

| Row | C | D | E | P |
|---|---|---|---|---|
| 1 | T | T | T | T |
| 2 | F | T | T | T |
| 3 | T | F | T | F |
| 4 | F | F | T | T |
| 5 | T | T | F | T |
| 6 | F | T | F | F |
| 7 | T | F | F | T |
| 8 | F | F | F | T |

**Figure 9 Determinacy Table for *P***

From Rows 1 and 3 of this table we can see that when α(C) = α(D) = T then the clause D determines the predicate P. Similarly, from Rows 6 and 8 when α (C) = α (E) = F we can also see that D determines P. From Rows 3 and 4 we see that clause C determines P. Finally, from Rows 3 and 7 we see that clause E determines P.

We will now look at some logic coverage models that generate test requirements to test determining clauses.

**Definition 12. Active Clause Coverage (ACC)** For each predicate $p \in P(G)$, and each clause $c \in C(p)$ which determines $p$ (under some α), there are two test requirements: (i) $tr(c)_{true} \in TR_{ACC}$ which implies that $c$ is reached and evaluates to *true*, and (ii) $tr(c)_{false} \in TR_{ACC}$ which implies that $c$ is reached and evaluates to *false*.

**Example:** In the example of Figure 9 Determinacy Table for *P* all three clauses C, D and E determine the predicate P. So, for active clause coverage of predicate P there are six test requirements

$$TR_{ACC} = \{\ tr(C)_{true},\ tr(C)_{false},\ tr(D)_{true},\ tr(D)_{false},\ tr(E)_{true},\ tr(E)_{false}\ \}.$$

This test requirement set $TR_{ACC}$ has some redundancy, and inspecting the table we can formulate the requirements more concisely as

$$TR_{ACC} = \{Row1, Row3, Row6, Row7, Row8\}.$$

In this slightly different notation, Row1 is shorthand for the test requirement C & D & E (c.f. Figure 8)

The definition of ACC can be seen as ambiguous. When considering a determining clause *c*, do the other clauses get the same assignment when *c* is *true* and *c* is *false*, or can they have different assignments? Furthermore, in practice, we may not be able to isolate the individual clauses so easily if they are not simple Boolean variables. Problems of *logical overlap* and *side-effects* (e.g. variable synonyms) between clauses can arise. Consider for example the predicate

$$p = !(x > 10) \mid (x > 0)$$

If the first clause is *true* then the second clause can never be *false*. Thus, not all combinations of clause values may be possible. This can be very difficult to detect in practice for the same reasons that make constraint satisfiability difficult to detect.

These observations lead to a fairly liberal approach to the problem of testing determining clauses.

**Definition 13. General Active Clause Coverage (GACC)** For each predicate $p \in P(G)$, and each clause $c \in C(p)$ which determines $p$, (under some $\alpha$), there are two test requirements: (i) $tr(c)_{true} \in TR_{GACC}$ which implies that $c$ is reached and evaluates to *true*, and (ii) $tr(c)_{false} \in TR_{GACC}$ which implies that $c$ is reached and evaluates to *false*. The values chosen for the other clauses $d \in C(p)$, $d \neq c$, need not be the same in both cases.

The important difference between ACC and GACC lies in the last sentence above.

**Example**: For GACC of clause D above there are 4 possible pairs of test requirements:
        (Row1, Row3), (Row1, Row6), (Row8, Row3), (Row8, Row6)

While GACC tests the determining clauses of a predicate, one problem with this definition is that GACC does not imply PC. This is because the values of the auxiliary clauses are not constrained in any way. Our goal was to unify PC with CC in an effective way, but we have lost PC. To see this, consider the predicate

$$p = a \leftrightarrow b$$

There exists an assignment $\alpha$ such that *a* determines *p*. The same holds for *b*. Now consider the following two test cases:

$tc_1$: (*a* = *true*, *b* = *true*) so *p* = *true*
$tc_2$: (*a* = *false*, b = *false*) so *p* = *true*

For the test suite { $tc_1, tc_2$ }, *p* never becomes false so 100% PC is not achieved, although 100% GACC is. The problem is that a correlation or side effect between the two clauses *a* and *b* has been deliberately introduced into the predicate *p*. However, in real life such correlations between clauses may be deeply hidden by the code structure and very difficult to identify.

Let us try to recover PC while maintaining good logical coverage of the determining clauses.

**Definition 14 Restricted Active Clause Coverage (RACC)** For each predicate $p \in P(G)$, and each clause $c \in C(P)$ which determines *p*, (under some α), there are two test requirements: (i) $tr(c)_{true} \in TR_{RACC}$ which implies that *c* is reached and evaluates to *true*, and (ii) $tr(c)_{false} \in TR_{RACC}$ which implies that *c* is reached and evaluates to *false*. The values chosen for the other clauses $d \in C(p)$, d ≠ c, must be the same in both cases.

**Example:** For RACC of clause D above there are 2 possible requirements sets (Row1, Row3), (Row6, Row8).

While superficially correct, the problem with RACC is that determination is not really a local property of the truth table, but is actually a global property of the entire code. This means that it may not be possible to keep the values for the auxiliary clauses the same.

To see this, consider for example the code snippet

```
x := y;
...
if (x>0 or y>0) then …
```

For predicate *p* = (x>0 or y>0) it looks like PC can be achieved by using RACC based on our determinacy analysis of (*a* | *b*) by taking *a* = (x > 0) and *b* = (y > 0). But this is misleading, since here x and y are effectively synonyms through the assignment x := y; and so 100% RACC cannot be achieved.  A different example of the same problem is given in (Ammann & Offutt, 2017), based on logical overlap between clauses.

It seems pointless to recommend a coverage model if it is not possible to reach 100% coverage according to that model. Even worse, it can be unclear what the maximum achievable coverage might be in any specific testing task. Our final definition relaxes RACC to what is always practically achievable.

**Definition 14. Correlated Active Clause Coverage (CACC)**
For each predicate $p \in P(G)$, and each clause $c \in C(P)$ which determines *p*, (under some α), there are two test requirements: (i) $tr(c)_{true} \in TR_{CACC}$ which implies that *c* is reached and evaluates to *true*, and (ii) $tr(c)_{false} \in TR_{CACC}$ which implies that *c* is reached and evaluates to *false*. The values chosen for the other clauses $d \in C(p)$, $d \neq c$, must cause *p* to be true in one case and false in the other.

**Example**: For CACC of clause D above there exist 2 possible test suites (Row1, Row3), (Row6, Row8).

In DO-178B, the Federal Aviation Authority (FAA) mandated a minimum level of logic coverage for ASIL level A (i.e. highest safety criticality level) avionic software which they termed "*Modified Condition Decision Coverage*" (MCDC). There was some confusion about this definition, based on the issues we have discussed above. The original definition (subsequently called "*Unique Cause MCDC*") was RACC. This is problematic, as we have seen. The revised definition (now called "*Masking MCDC*") is CACC, which permits 100% achievable coverage.

## Data Flow Coverage

We now consider a final class of coverage models, which are the data flow coverage models. Although data flow coverage also yields test requirements that are paths in a condensation graph, these are intended to be orthogonal to control flow requirements. If we accept the "definition":

$$\text{algorithms + data structures = programs}^{12}$$

then data flow coverage is intended to test errors concerning data structures. This typically means incorrect definition or use of some data operations. The effect of this will usually be seen in the values of the program variables, while the control flow might be completely unaffected. We begin with some technical definitions.

**Definition 15.** Let *G* be a condensation graph.

1. A *definition* of a variable *v* in *G* is any statement in *G* that writes to *v*.
2. A *use* of *v* is any statement that reads *v*.
3. A path $p = (n_1, ...., n_k)$ in G from a node $n_1$ to a node $n_k$ is *def-clear* for *v* if for each $1 < j < k$, the node $n_j$ has no statements which write to *v*.

**Definition 16.** Let *G* be a condensation graph. A *du-path* w.r.t. *v* is a simple path
$$p = (n_1, ..., n_k)$$
such that:

1. A statement in $n_1$ writes to *v*.

---

[12] This definition was first proposed in (Wirth, 1976) and still seems relevant to software testing today.

2. Path *p* is def-clear for *v*.
3. A statement in $n_k$ reads *v*.

We let $du(n, v)$ denote the set of all du-paths wrt *v* starting at node *n*. We let $du(m, n, v)$ denote the set of all du-paths wrt *v* starting at node *m* and ending at node *n*.

Recall that $p = (n_1, ..., n_k)$ is simple when *p* has no repetitions of nodes other than (possibly) the first and last.

**Definition 17. All-defs Coverage (ADC)** For each def-path set $S = du(n, v)$ the test requirement set $TR_{ADC}$ contains at least one path $d \in S$.

**Definition 18. All-uses Coverage (AUC)** For each def-pair set $S = du(m, n, v)$ the test requirement set $TR_{AUC}$ contains at least one path $d \in S$.

**Definition 19. All-du-paths Coverage (ADUPC)** For each def-pair set $S = du(m, n, v)$ the test requirement set $TR_{ADUPC}$ contains every path $d \in S$.

Figure 10 Data Flow Coverage Example shows an example condensation graph *G* and the corresponding test requirement set generated by each of the three data flow coverage models.

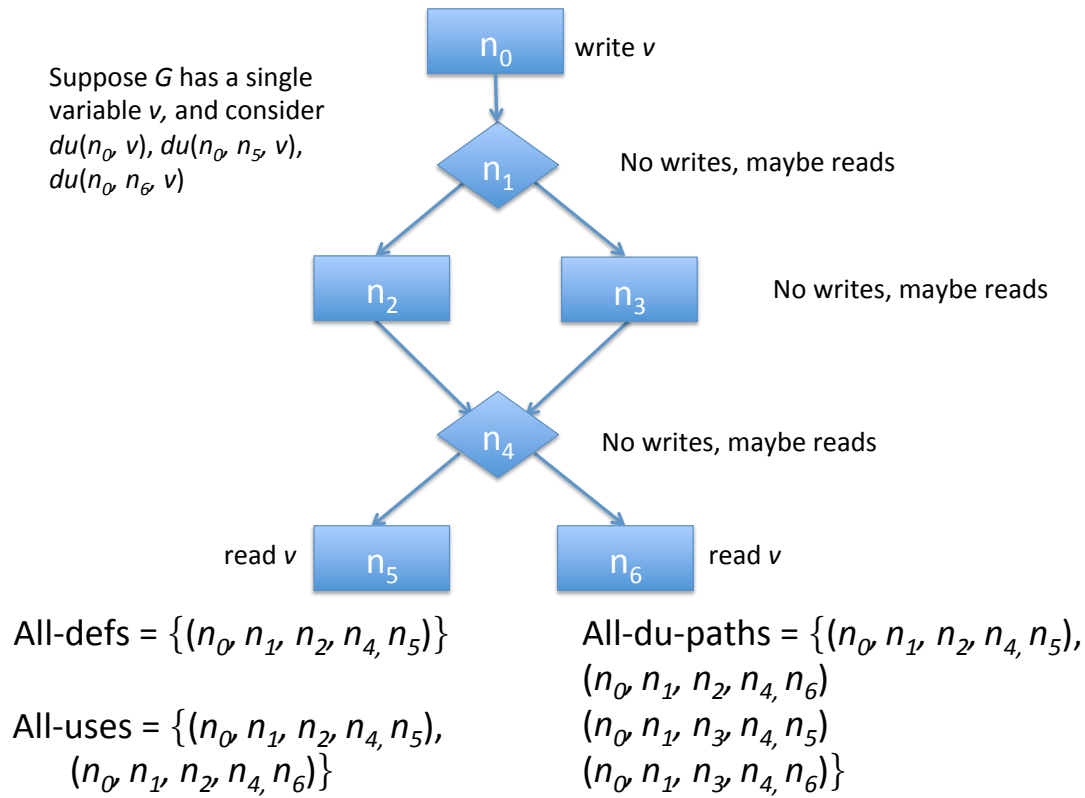

**Figure 10 Data Flow Coverage Example**

## Constructing Test Cases from Coverage Requirements

In this section we consider the problem of constructing a test case *tc* that satisfies a given test coverage requirement *tr*. For some reason, this subject is frequently omitted in introductory texts to software testing. Perhaps there is an assumption that the problem is relatively straightforward and easy to solve. For simple programs this can be true. However, in general, test case construction is a difficult, error prone and time-consuming task if it must be done manually. As we have already mentioned, on theoretical grounds there is no general algorithm for test case construction that solves the problem in every situation. This is due to the algorithmic undecidability of a great many constraint solving problems, starting with the famous Halting Problem (see (Sipser, 2012)). Even those constraint solving problems which are algorithmically solvable (such as the Boolean SAT problem), often have a prohibitive time complexity for large problem instances.

In this section we will present a systematic approach to test case construction from a given path requirement *tr* by *symbolic evaluation* of the path. This means we combine the code information found along the path (assignments and Boolean conditions) into a single Boolean constraint that characterizes the test cases satisfying that path requirement. By learning a simple set of rules for assignments and conditions, and by combining these rules with common algebraic laws about programming language predicates and operators (such as equations, inequalities and the laws of Boolean algebra) it is possible to rigorously derive valid test case constraints. The goal is to correctly derive a constraint C, in such a way that:

(i)  we can assert with confidence that a test coverage requirement *tr* is *satisfiable* by some test case *tc*, and can correctly construct a satisfying test case *tc*, either manually or with the help of an algorithm or tool.
(ii) we can detect if a test requirement *tr* is *unsatisfiable*. In this case, there is typically a code component that is unreachable or "dead code".

We will illustrate the process of symbolic evaluation to derive a test case constraint from a path requirement by means of a simple concrete example. Hopefully this example is sufficiently generic that the reader can apply the same principles to other examples.

Consider the following simple code example.

```
x = 2*x;
y = x+y;

while (y+x > 100) {

    x = x +1;
    if (x mod 2 == 0){ y = y-1; } else { y = y-2; }
```

```
};
return y;
```

The condensation graph for this code is given in Figure 11 Condensation graph.

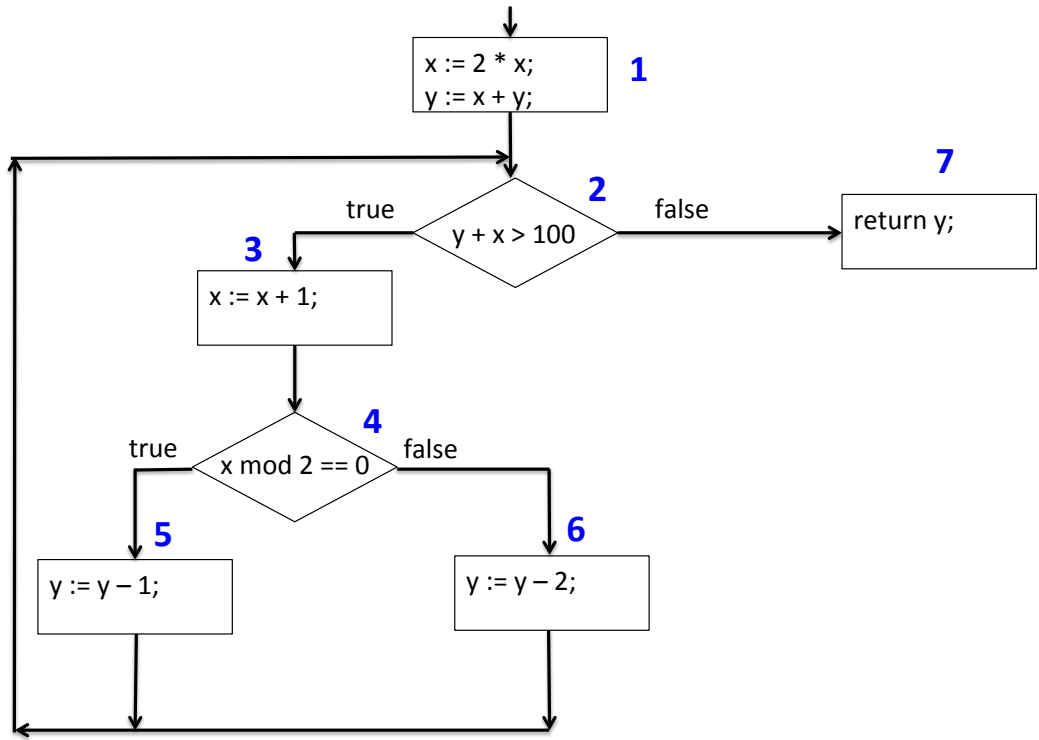

**Figure 11 Condensation graph**

This graph clearly shows typical CFG features such as a loop and a branch.

Consider the concrete test requirement *tr* which is the path

    *tr* = 1,2,3,4,6,2,7

This path *tr* traverses the loop body exactly once.

In general, for any test requirement *tr* = $v_1, ..., v_k$, we wish to be able to find a test case *tc*, (i.e. an assignment of values to the SUT input variables) that traverses *all* of the nodes $v_1, ..., v_k$. We would really like a general method that solves the problem of finding *tc* for any condensation graph and any path requirement.

Our approach will be to derive a logical constraint $C(x_1, ..., x_n)$ on the input variables $x_1, ..., x_n$ of the SUT (for the above SUT these are x, y) such that:

(i) every solution[ii] α of C($x_1$,..., $x_n$) leads SUT execution down the path *tr*, and

(ii) if α is a test case that leads execution down the path tr then α is a solution of the constraint C($x_1$,..., $x_n$).

So C($x_1$,..., $x_n$) will exactly characterize all the test cases satisfying *tr*, and if the set of solutions of C($x_1$,..., $x_n$) is empty, i.e. C is unsatisfiable then we can be certain that *tr* is a dead or unreachable path.

What makes the problem of constructing C($x_1$,..., $x_n$) difficult is that although the Boolean conditions of the original code are the key to finding C, the presence of assignment statements (such as those in node 1 above) has a modifying effect on the actual code conditions. To combine the relevant assignments and Boolean conditions of *tr* into a single constraint, we must traverse *tr* as a path from start to finish, and collect and integrate the assignments and Boolean conditions along the way. Often it is very useful to simplify the constraint C using common mathematical laws, such as the laws of Boolean algebra, after each path step. In this way we may more easily find test cases as solutions.

In Table 1 Constraint table for the path tr = 1,2,3,4,6,2,7 below we present a *constraint table* for the path *tr* = 1,2,3,4,6,2,7.

This table consists of one line for each code instruction along the path. We number the associated graph node of the instruction in the left-hand column. We break each assignment block down into its individual assignment statements, to make each step in the construction of C smaller and simpler. In the middle column, we show the actual code instruction along the path *tr*. In the right-hand column of the table, we show the *current constraint* arising from either:
  (i) executing the assignment on the current line, or
  (ii) evaluating the Boolean condition on the current line to either *true* or *false*, depending on which next node we progress to.

Notice in the constraint column that each code variable can appear as itself, e.g. **x** or in primed form e.g. **x'**. The unprimed variable e.g. **x** stands for the value of that variable at the *start of code execution*. An unprimed variable is a variable we must ultimately find a value for, in order to build a test case. The primed variable e.g. **x'** stands for the value of **x** *after execution of the instruction* in the middle column. So, the constraint on each line expresses a relationship between the values of program variables *initially* and their values *currently* on each line.

For example, in line 1, the code instruction in column 2 from node 1 is **x := 2 \* x**. So not surprisingly the relationship between **x** before this instruction and **x** after this instruction (i.e. **x'**) in column 3 is:

$$x' = 2 * x$$

Line 2 has a second assignment instruction **y := x + y** from node 1. To understand its effect on the constraint C in column 3, we must replace the variable **x** in the rhs expression **x + y** by its current value immediately before the assignment, which is **x'**. Now according to line 1, **x' = 2 * x.** Substituting **x** with **2 * x** in **x + y** gives

$$y' = 2*x + y.$$

In line 3 we encounter the loop guard **y + x > 100**. In order to proceed to node 3 we must make this Boolean condition *true*. Now **y + x > 100** will be true precisely when the current value of **x** plus the current value of **y** is greater than **100**. From lines 1 and 2 we can see these current values **x'** and **y'**. They are again substituted into the code line **y + x > 100** giving the constraint **((2*x)+y) + 2*x > 100** on line 3 in the right hand column. In this way, making assignment substitutions and condition substitutions using the current values of variables from previous lines, we can derive a constraint for every code line.

In particular we eventually see that to arrive at node 2 for the second time we must satisfy the constraint

$$4x + y = 101$$

This shows that the only test cases which lead down *tr* are assignments α(**x**), α(**y**) for **x** and **y** which satisfy the linear equation **4α(x) + α(y) = 101**. For example, let α(**x**) = 25 and α(**y**) = 1 then the constraint is satisfied. Manually tracing the code execution for these input values should confirm our calculations.
.

| Graph Node | Instruction | Constraint |
|---|---|---|
| 1 | x := 2 * x | x' = 2 * x |
| 1 | y := x + y | y' = 2*x + y |
| 2 | y + x > 100 | ((2*x)+y) + 2*x > 100 |
| 3 | x := x + 1 | x' = 2*x + 1 |
| 4 | x mod 2 != 0 | ((2*x)+y) + 2*x > 100 & (2*x + 1 mod 2 != 0) <br><br> ≡ <br><br> ((2*x)+y) + 2*x > 100 & TRUE <br><br> ≡ <br><br> ((2*x)+y) + 2*x > 100 |
| 6 | y := y − 2 | y' = (2*x + y) - 2 |
| 2 | y + x <= 100 | ((2*x + y) − 2) + (2*x + 1) <= 100 & <br> ((2*x)+y) + 2*x > 100 & <br> (2*x + 1 mod 2 != 0) <br><br> ≡ <br><br> ((2*x + y) − 2) + (2*x + 1) <= 100 & |

| | | ((2*x)+y) + 2*x > 100 & TRUE ≡ 4x + y > 100 & 4x + y <= 101 ≡ 4x + y = 101 |
|---|---|---|
| 7 | return y | 4x + y = 101 |

Table 1 Constraint table for the path tr = 1,2,3,4,6,2,7

The constraint derived on test cases is by no means always as simple as a linear equation (which always has a solution). Table 2 constraint table for the path tr = 1,2,3,4,5,2,7 gives a similar analysis for the path *tr* = **1,2,3,4,5,2,7** in the same condensation graph. This time we see from line 5 that the constraint C(x, y) is always false, i.e. unsatisfiable. This means there is no test case that leads to execution of *tr*. The code at node 5 is therefore dead or unreachable code.

| Graph Node | Instruction | Constraint |
|---|---|---|
| 1 | x := 2 * x | x' = 2 * x |
| 1 | y := x + y | y' = 2*x + y |
| 2 | y + x > 100 | ((2*x)+y) + 2*x > 100 |
| 3 | x := x + 1 | x' = 2*x + 1 |
| 4 | x mod 2 == 0 | ((2*x)+y) + 2*x > 100 & (2*x + 1 mod 2 = 0) ≡ ((2*x)+y) + 2*x > 100 & FALSE ≡ FALSE |
| 5 | y := y – 1 | **Unreachable along this path** |
| 2 | y + x <= 100 | **Unreachable along this path** |
| 7 | return y | **Unreachable along this path** |

Table 2 constraint table for the path tr = 1,2,3,4,5,2,7

## Conclusions
To be added after review.

## Acknowledgements
The author gratefully acknowledges financial support from the EU ITEA3 TESTOMAT Project 16032.

---

[i]

[ii] The notation $C(x_1, ..., x_n)$ means that C is a logical formula in which only the variables $x_1, ..., x_n$ appear. A solution of $C(x_1, ..., x_n)$ is an assignment of values to $x_1, ..., x_n$ that makes $C(x_1, ..., x_n)$ true.